\documentclass[12pt]{article}

\usepackage{float}

\usepackage[date=year,style=numeric,giveninits=true,sorting=none]{biblatex}
\AtEveryBibitem{\clearlist{language} \clearlist{location} }
\renewbibmacro{in:}{}

\usepackage[dvipdfm,margin=1.1in]{geometry}
\usepackage{amsfonts,mathtools,amsthm,bm}
\usepackage{subfiles}
\usepackage{subcaption}
\usepackage[dvipdfmx]{graphicx}
\usepackage{authblk}
\DeclareMathOperator{\sgn}{sgn}
\usepackage[dvipdfmx]{color}
\usepackage[dvipdfmx]{hyperref}
\hypersetup{
setpagesize=false,
 colorlinks=true,%
 linkcolor=magenta,
 citecolor=blue,
}
\newtheorem{theorem}{Theorem}[section]

\newtheorem{corollary}[theorem]{Corollary}
\newtheorem{lemma}[theorem]{Lemma}
\newtheorem{proposition}[theorem]{Proposition}

\title{\textbf{Localization in quantum walks with periodically arranged coin matrices} }
\author[]{Chusei Kiumi}
\affil[]{\footnotesize Mathematical Science Unit, 
Graduate School of Engineering Science,\protect\\
Yokohama National University,\protect\\
Hodogaya, Yokohama, 240-8501, Japan}
\date{\empty}

\begin{document}
\maketitle
\vspace{-1.2cm}
\begin{abstract}
There is a property called localization, which is essential for applications of quantum walks. From a mathematical point of view, the occurrence of localization is known to be equivalent to the existence of eigenvalues of the time evolution operators, which are defined by coin matrices. A previous study proposed an approach to the eigenvalue problem for space-inhomogeneous models using transfer matrices. However, the approach was restricted to models whose coin matrices are the same in positions sufficiently far to the left and right, respectively. This study shows that the method can be applied to extended models with periodically arranged coin matrices. Moreover, we investigate localization by performing the eigenvalue analysis and deriving their time-averaged limit distribution.
\end{abstract}

\section{Introduction}
The study of quantum walks, which began in the early 2000s \cite{Ambainis2001-os,Konno2002-sh}, has spread and attracted much attention, especially for its applications in quantum information \cite{Childs2004-xa,Childs2009-er,Lovett2010-en,Childs2013-zk,Falcao2021-pb,Machida2015-oa}. One of the most characteristic properties of the quantum walk is localization, an essential property for manipulating particles. Numerical and theoretical analyses have been actively conducted, and we are particularly interested in the mathematical analysis of localization
\cite{Cantero2012-yk,Endo2020-or,Inui2005-fr,Machida2015-oa,Segawa2016-qu,Wojcik2012-kr}. It has been known that localization occurs in various quantum walk models. This study focuses on one-dimensional two-state quantum walks, which are considered the fundamental discrete-time quantum walks.

From a mathematical point of view, the investigation of localization can be regarded as an eigenvalue problem. This is because the occurrence of localization is equivalent to the existence of eigenvalues of the time evolution operator, and the corresponding eigenvectors are related to how likely the walker localizes
\cite{Segawa2016-qu,Kiumi2021-dp}. In a previous study \cite{Kiumi2021-yg}, an eigenvalue analysis method using the transfer matrix was proposed. The eigenvalue analysis was performed for two-phase quantum walks with one defect, including a one-defect model where the time evolution differs at the origin and a two-phase model where the time evolution differs in the negative and positive parts, respectively. The localization phenomenon in the one-defect model has been used in quantum search algorithms \cite{Shenvi2003-jw,Childs2004-xa,Ambainis2005-ha}, and the relationship between topological insulators and localization in the two-phase quantum walk has attracted much attention \cite{Kitagawa2010-su,Endo2015-db}. Several other studies also used transfer matrices for deriving stationary measures \cite{Kawai2018-ry,Kawai2017-fn}. Furthermore, a previous study \cite{Kiumi2021-dp,Kiumi2021-rm} showed that the method could be applied to a more general model with a finite number of defects, which satisfies the following conditions:
\[
C_x =
\begin{cases}
C_{+\infty} ,\quad & x\in [ x_{+} ,\infty ),
\\
C_{-\infty} ,\quad & x\in ( -\infty ,x_{-}],
\end{cases}
\]
where $x$ and $x_+>0,x_-<0$ are integers, and $C_x$ denotes the coin matrix determining the time evolution in the position $x$. Models with periodic coin matrices have also been actively studied \cite{linden2009inhomogeneous,ahmad2020one,Shikano2010}. In particular, the model with self-duality studied in \cite{Shikano2010} is inspired by the well-studied Aubry-André model \cite{Aubry1980-ys}, and the Fourier transform method was applied. However, our approach can extend the discussion to models that cannot be handled by the Fourier transform method and simplify the eigenvalue problem since we only need to deal with 2 $\times$ 2 (transfer) matrices instead of the large matrix. In this study, we consider a two-phase periodic model with a finite number of defects that includes all of the above important models; the model has $n_-$ and $n_+$ different coin matrices arranged periodically in positions $x<x_-$ and $x\geq x_+$, respectively. 

This paper is organized as follows. In Section \ref{sec:2}, we define our model with periodically arranged coin matrices and its transfer matrix. This section also shows that the eigenvalue analysis via transfer matrix is applicable for the model. The necessary and sufficient condition for the eigenvalue problem is given in Theorem \ref{Theorem Ker}. Also, further discussion about the time-averaged limit distribution using eigenvalues and eigenvectors is provided. With the main theorem, Section \ref{sec:3} focuses on analyzing concrete eigenvalues for more specific models, which can be seen as natural generalizations of homogeneous, one-defect, and two-phase models to periodic models.

\section{Definitions and method}
\label{sec:2}
Firstly, we introduce two-state quantum walks on the integer lattice $\mathbb{Z}$.
The Hilbert space is given as $\mathcal{H}$ defined by
\[
\mathcal{H}=\ell^2(\mathbb{Z} ; \mathbb{C}^2) =
\left\{
\Psi : \mathbb{Z} \to \mathbb{C}^2\ \middle\vert\ \sum_{x\in\mathbb{Z}}\|\Psi(x)\|_{\mathbb{C}^2}^2 < \infty
\right\},
\]
where $\mathbb{C}$ denotes the set of complex numbers. For $\Psi\in\mathcal{H}$, we write $\Psi(x)=\begin{bmatrix}\Psi_L(x) & \Psi_R(x)\end{bmatrix}^T\in\mathbb{C}^2$, where $T$ is the transpose operator.  The time evolution operator $U$ is defined by the product of a coin operator $C$ on $\mathcal{H}$ and a shift operator $S$ on $\mathcal{H}$. Here, $C$ and $S$ are defined by  
 \begin{align*}
	(C\Psi)(x)
	=C_x\Psi(x),\qquad
	(S\Psi)(x)=\begin{bmatrix}
	\Psi_L(x+1)	
	\\ \Psi_R(x-1)
	\end{bmatrix},
\end{align*}
	where $\{C_x\}_{x\in\mathbb{Z}}$ is a sequence of $2\times2$ unitary matrices called coin matrices.
 We define $x_+,x_-\in \mathbb{Z}$ where $x_+\geq0,\ x_-\leq0$, then the periodicities of the model are defined by $n_+\in\mathbb{Z}_{>0}$ for positions $x\geq x_+$ and  $n_-\in\mathbb{Z}_{>0}$ for positions $x<x_-$, where $\mathbb{Z}_{>0}$ is the set of positive integers. For $x\in \mathbb{Z}$, we define $r_{x}^{\pm } \in \{0,\dots ,n_{\pm } -1\}$ by remainder of $x-x_{\pm }$ divided by $n_\pm\in\mathbb{Z}_{>0}$, that is, 
\[
r_{x}^{\pm } =x-x_{\pm } \ (\bmod \ n_{\pm }).
\]
 We treat the model whose coin matrices satisfy the following:
\[
C_{x} =\begin{cases}
C_{r_{x}^{+}}^{+} , & x\in [ x_{+} ,\infty ) ,\ \\
C_{x} & x\in [ x_{-} ,x_{+}) ,\\
C_{r_{x}^{-}}^{-} , & x\in ( -\infty ,x_{-}) .
\end{cases}
\]
The model has a finite number of defects in $x\in[x_-,x_+)$ and periodic coin matrices for each of $x\geq x_+$ and $x<x_-$. Here, we write $C_x,C^{\pm}_{k}$ as follows:
    \begin{align*}
C_{x} =e^{i\Delta _{x}}\left[\begin{array}{ c c }
\alpha _{x} & \beta _{x}\\
-\overline{\beta _{x}} & \overline{\alpha _{x}}
\end{array}\right] ,\ \ C_{k}^{\pm } =e^{i\Delta _{k}^{\pm }}\left[\begin{array}{ c c }
\alpha _{k}^{\pm } & \beta _{k}^{\pm }\\
-\overline{\beta _{k}^{\pm }} & \overline{\alpha _{k}^{\pm }}
\end{array}\right],
	\end{align*}
	with $ \alpha _{x} ,\beta _{x} ,\alpha _{k}^{\pm } ,\beta _{k}^{\pm } \in \mathbb{C},\ \Delta _{x} ,\Delta _{k}^{\pm } \in [0,2\pi ),\ | \alpha _{x}| ^{2} +| \beta _{x}| ^{2} =\left| \alpha _{k}^{\pm }\right| ^{2} +\left| \beta _{k}^{\pm }\right| ^{2} =1,$ and $\alpha _{x} ,\alpha _{k}^{\pm } \neq 0$ for $ x\in \mathbb{Z} ,\ k\in \{0,\dotsc ,n_{\pm } -1\}$.
	
	We let $\Psi_0\in\mathcal{H}\ (\|\Psi_0\|_{\mathcal{H}}^2=1)$ denote the initial state of the model. Then, the probability distribution at time $t\in\mathbb{Z}_{\geq 0}$ is defined by  \[\mu_t^{(\Psi_0)}(x)=\|(U^t\Psi_0)(x)\|_{\mathbb{C}^2}^2,\]
	 where $\mathbb{Z}_{\geq 0}$ is the set of non-negative integers. Here, we say that the quantum walk model exhibits localization if there exists an initial state $\Psi_0\in\mathcal{H}$ and a position $x_0\in\mathbb{Z}$ satisfying \[\limsup_{t\to\infty}\mu^{(\Psi_0)}_t(x_0)>0.\] 
	 As a well-known fact that the quantum walk model exhibits localization if and only if the time evolution operator $U$ has an eigenvalue \cite{Segawa2016-qu}, which means that there exists $\lambda\in[0, 2\pi)$ and $\Psi\in\mathcal{H}\setminus\{\mathbf{0}\}$ such that
\begin{align*}
	U\Psi=e^{i\lambda}\Psi.
	\end{align*}
 Let $\sigma(U)$ denotes the set of eigenvalues of $U$ henceforward. Subsequently, let $J$ be a unitary operator on $\mathcal{H}$ defined by
\begin{align*}
	    (J\Psi)(x)=
	    \begin{bmatrix}
	    \Psi_L(x-1) \\ \Psi_R(x)
	    \end{bmatrix}
	\end{align*}
	for $x\in\mathbb{Z}$. Here the inverse of $J$ is given as
\begin{align*}
	   	(J^{-1}\Psi)(x)=
	    \begin{bmatrix}
	    \Psi_L(x+1) \\ \Psi_R(x)
	    \end{bmatrix}.
	\end{align*}
 Furthermore, for $\lambda\in[0,2\pi)$, $x\in\mathbb{Z}$ and $k\in \{0,\dotsc ,n_{\pm } -1\}$, we introduce the transfer matrix $T_x(\lambda),\ T^{\pm}_{k}(\lambda)$ as followings:
	\[
	T_{x}( \lambda ) =\frac{1}{\alpha _{x}}\left[\begin{array}{ c c }
e^{i( \lambda -\Delta _{x})} & -\beta _{x}\\
-\overline{\beta _{x}} & e^{-i( \lambda -\Delta _{x})}
\end{array}\right] ,\ \ T_{k}^{\pm }( \lambda ) =\frac{1}{\alpha _{k}^{\pm }}\left[\begin{array}{ c c }
e^{i\left( \lambda -\Delta _{k}^{\pm }\right)} & -\beta _{k}^{\pm }\\
-\overline{\beta _{k}^{\pm }} & e^{-i\left( \lambda -\Delta _{k}^{\pm }\right)}
\end{array}\right] .
\]
	 We abbreviate the transfer matrix $T_x(\lambda)$ as $T_x$ and $T^{\pm}_{k}(\lambda)$ as $T^{\pm}_{k}$ henceforward. Here, $\Psi\in\mathcal{H}$ satisfies $(U-e^{i\lambda})\Psi = 0$ if and only if $\Psi$ satisfies the following equation for all $x\in\mathbb{Z}$ :
	\begin{align}
	\label{eq:relation}
	    (J\Psi)(x+1) = T_x(J\Psi)(x).
	\end{align}
	For more details, see \cite{Kiumi2021-yg}.
	In this paper, we define $\prod$ notation for the products of matrices by the following:
	\[
	\prod _{i=k}^{n} A_{i} =\begin{cases}
A_{n} \cdots A_{k+1} A_{k}, & n\geq k,\\
1, & n< k.
\end{cases}
\]
For $\lambda\in[0,2\pi)$ and $\varphi\in\mathbb{C}^2$, we define $\tilde\Psi : \mathbb{Z}\to \mathbb{C}^2$ as follows:
\begin{align*}
\tilde{\Psi } (x) 
&=\left\{\begin{array}{ l l }
T_{x-1} T_{x-2} \cdots T_{1} T_{0} \varphi , & x >0,\\
\varphi , & x=0,\\
T_{x}^{-1} T_{x+1}^{-1} \cdots T_{-2}^{-1} T_{-1}^{-1} \varphi , & x< 0.
\end{array}\right. \nonumber
\\
&=\begin{cases}
\displaystyle\prod _{i=0}^{r_{x}^{+} -1} T_{i}^{+ }\left(\prod _{i=0}^{n_{+} -1} T_{i}^{+ }\right)^{m_{x}^{+}} T_{+} \varphi , & x >x_{+} ,\\
\displaystyle\prod _{i=0}^{x-1} T_{i} \varphi , & 0< x\leq x_{+} ,\\
\varphi , & x=0,\\
\displaystyle\prod _{i=1}^{| x| } T_{-i}^{-1} \varphi , & x_{-} \leq x< 0,\\
\displaystyle\left(\prod _{i=r_{x}^{-}}^{n_{-} -1} T_{i}^{- }\right)^{-1} \ \left(\prod _{i=0}^{n_{-} -1} T_{i}^{- }\right)^{-m_{x}^{-}} T_{-} \varphi , & x< x_{-} ,
\end{cases}
\end{align*}
where $T_{+} =\prod _{i=0}^{x_{+} -1} T_{i},\  T_{-} =\prod _{i=1}^{| x_{-}| } T_{-i}^{-1}$ and $m_{x}^{+} =\frac{x-x_+-r_{x}^{+}}{n_{+}} \in \mathbb{Z}_{\geq 0} ,\ \  m_{x}^{-} =\frac{\left| x-x_- -r_{x}^{-}+n_{-}\right| }{n_{-}} \in \mathbb{Z}_{\geq 0}.$
This can be rewritten as
	\begin{align}
	\label{eq:tilde_psi}
\tilde{\Psi } (x) 
=\begin{cases}
\displaystyle\left(\tilde{T}_{r_{x}^{+}}^{+}\right)^{m_{x}^{+}}\prod _{i=0}^{r_{x}^{+} -1} T_{i}^{+ } T_{+} \varphi , & x >x_{+} ,\\
\displaystyle\prod _{i=0}^{x-1} T_{i} \varphi , & 0< x\leq x_{+} ,\\
\varphi , & x=0,\\
\displaystyle\prod _{i=1}^{| x| } T_{-i}^{-1} \varphi , & x_{-} \leq x< 0,\\
\displaystyle\left(\tilde{T}_{r_{x}^{-}}^{- }\right)^{-m_{x}^{-}} \ \left(\prod _{i=r_{x}^{-}}^{n_{-} -1} T_{i}^{- }\right)^{-1} T_{-} \varphi , & x< x_{-} ,
\end{cases}
\end{align}
with
\[\tilde{T}_{k}^{\pm  } =\prod _{i=0}^{k-1} T_{i}^{\pm }\prod _{i=k}^{n_{\pm } -1} T_{i}^{\pm  } ,
\]
for $k\in \{0,\dotsc ,n_{\pm } -1\}.$ Here, $\tilde\Psi:\mathbb{Z}\rightarrow\mathbb{C}^2$ is a map constructed by transfer matrices, where $\Psi=J^{-1}\tilde\Psi$ satisfies equation (\ref{eq:relation}) (but not necessarily satisfies $\sum_{x\in\mathbb{Z}}\|\Psi(x)\|_{\mathbb{C}^2}^2 < \infty$). For $\lambda\in[0,2\pi)$, $\tilde\Psi$ has $\varphi\in\mathbb{C}^2$ as a parameter. We let $W_{\lambda}$ be a set of all possible $\tilde\Psi$ obtained by varying $\varphi$:
\[W_{\lambda}=\left\{\tilde\Psi:\mathbb{Z}\rightarrow\mathbb{C}^2\ \middle|\  \tilde\Psi(x)\text{ is given by }  (\ref{eq:tilde_psi}),\  \varphi\in\mathbb{C}^2\right\}.\]
	\begin{corollary}
	\label{CORO_A}
 Let $\lambda\in [0,2\pi)$, $e^{i\lambda}\in\sigma(U)$ if and only if there exists $\tilde\Psi\in W_{\lambda} \setminus\{\mathbf{0}\}$ such that $\tilde\Psi\in\mathcal{H}$, and associated eigenvector of $e^{i\lambda}$ becomes $J^{-1}\tilde\Psi$.
    \end{corollary}
Note that $\Psi=J^{-1}\tilde\Psi$ where $\tilde\Psi\in W_{\lambda}\setminus\{\mathbf{0}\}$ but not necessarily $\tilde\Psi\in\mathcal{H}$ is the stationary measure of the quantum walk studied in \cite{Endo2014-jx,Wang2015-oy,Kawai2017-fn,Kawai2018-ry,Endo2019-ie}. We define sign function for real numbers $u$ as follows:
\[
\sgn( u) =\begin{cases}
1, & u >0,\\
0, & u=0,\\
-1, & u< 0.
\end{cases}
\]
Then, two eigenvalues of $ \tilde{T}_{k}^{\pm  }\ (k\in \{0,\dotsc ,n_{\pm } -1\})$ can be written as $ \zeta _{\pm }^{ >} ,\zeta _{\pm  }^{< }$ expressed as below:
\begin{align*}
 & \zeta _{\pm }^{ >} =\frac{A^{\pm } +\sgn\left( A^{\pm }\right)\sqrt{\left( A^{\pm }\right)^{2} -\left| \alpha _{0}^{\pm }\right| ^{2}\left| \alpha _{1}^{\pm }\right| ^{2} \cdots \left| \alpha _{n_{\pm } -1}^{\pm }\right| ^{2}}}{\alpha _{0}^{\pm } \alpha _{1}^{\pm } \cdots \alpha _{n_{\pm } -1}^{\pm }} ,\\
 & \zeta _{\pm }^{< } =\frac{A^{\pm } -\sgn\left( A^{\pm }\right)\sqrt{\left( A^{\pm }\right)^{2} -\left| \alpha _{0}^{\pm }\right| ^{2}\left| \alpha _{1}^{\pm }\right| ^{2} \cdots \left| \alpha _{n_{\pm } -1}^{\pm }\right| ^{2}}}{\alpha _{0}^{\pm } \alpha _{1}^{\pm } \cdots \alpha _{n_{\pm } -1}^{\pm }} ,
\end{align*}
where $\displaystyle A^{\pm } =\frac{1}{2} \alpha _{0}^{\pm } \alpha _{1}^{\pm } \cdots \alpha _{n_{\pm } -1}^{\pm }\mathrm{tr}\left(\prod _{i=0}^{n_{\pm } -1} T_{i}^{\pm }\right)$. Note that the eigenvalues are independent of $k$.
\begin{corollary}
\label{cor:mat}
 $A^{\pm }$ is a real number.
\proof{
Let
\[
M=\left\{\left[\begin{array}{ c c }
a & b\\
\overline{b} & \overline{a}
\end{array}\right]\middle| \ a,b\in \mathbb{C}\right\}.\]
Here, $m_1m_2\in M$ holds for all $m_1,m_2\in M$, which implies that $m_1m_2\cdots m_n\in M\ (n>1)$ holds for all $m_1,m_2,\dots, m_n\in M$.
Also, $\alpha^{\pm}_k T_{k}^{\pm }\in M\ (k\in \{0,\dotsc ,n_{\pm } -1\})$ holds, thus we have $\alpha _{0}^{\pm } \alpha _{1}^{\pm } \cdots \alpha _{n_{\pm } -1}^{\pm }\prod _{i=0}^{n_{\pm } -1} T_{i}^{\pm }\in M$ and $A^{\pm }$ becomes a real number.
}

\end{corollary}
 Subsequently, since $\displaystyle \left| \det T_{k}^{\pm }\right| =\left| \frac{\overline{\alpha _{k}^{\pm  }}}{\alpha _{k}^{\pm  }}\right| =1$ for $k\in \{0,\dotsc ,n_{\pm }-1\},$
\[
\left| \det\tilde{T}_{k}^{\pm }\right| =\left| \frac{\overline{\alpha _{0}^{\pm }}\overline{\alpha _{2}^{\pm }} \cdots \overline{\alpha _{n_{\pm } -1}^{\pm }}}{\alpha _{0}^{\pm } \alpha _{2}^{\pm } \cdots \alpha _{n_{\pm } -1}^{\pm }}\right| =1\]
 holds, which also means $|\zeta _{\pm  }^> ||\zeta _{\pm  }^< |=1$. From Corollary \ref{cor:mat}, we know that $|\zeta _{\pm  }^{ >} |\geq 1$ and $|\zeta _{\pm  }^{< } |\leq 1$ hold, and we have the main theorem.
\begin{theorem}
\label{Theorem Ker}
For $\lambda\in[0,2\pi)$, $e^{i\lambda} \in \sigma( U) $ if and only if the following two statements hold:
\begin{align*}
    &1.\ \left( A^{\pm }\right)^{2} -\left| \alpha _{0}^{\pm }\right| ^{2}\left| \alpha _{1}^{\pm }\right| ^{2} \cdots \left| \alpha _{n_{\pm } -1}^{\pm }\right| ^{2}  >0,\\
    &2.\ \ker\left(\left(\prod _{i=0}^{n_{+} -1} T_{i}^{+} -\zeta _{+}^{< }\right) T_{+}\right) \cap \ker\left(\left(\prod _{i=0}^{n_{-} -1} T_{i}^{-} -\zeta _{-}^{ >}\right) T_{-}\right) \neq \{\mathbf{0}\}.
\end{align*}

\begin{proof}
From Corollary \ref{CORO_A}, $e^{i\lambda} \in \sigma( U)$ if and only if there exists $\varphi\in\mathbb{C}^2$ such that $\tilde\Psi\in W_{\lambda} \setminus\{\mathbf{0}\}$ given by $\varphi$ satisfies $\sum_{x\in\mathbb{Z}}\|\tilde\Psi(x)\|_{\mathbb{C}^2}^2<\infty$. If $\left( A^{\pm  }\right)^{2} -\left| \alpha _{0}^{\pm  }\right| ^{2}\left| \alpha _{1}^{\pm  }\right| ^{2} \cdots \left| \alpha _{n_{\pm } -1}^{\pm  }\right| ^{2} \leq0$, both $|\zeta ^<_{\pm}|$ and $|\zeta^>_{\pm}|$ become 1. Since $\tilde{\Psi}$ is given as (\ref{eq:tilde_psi}), we have $\sum_{x\in\mathbb{Z}}\|\tilde\Psi(x)\|_{\mathbb{C}^2}^2=\infty $ for all $\tilde\Psi\in W_\lambda\setminus\{\mathbf{0}\}$. Therefore, the first condition is necessary for $e^{i\lambda}\in\sigma(U) $. Next, we assume that $\left( A^{\pm  }\right)^{2} -\left| \alpha _{0}^{\pm  }\right| ^{2}\left| \alpha _{1}^{\pm  }\right| ^{2} \cdots \left| \alpha _{n_{\pm } -1}^{\pm  }\right| ^{2}  >0$, then $|\zeta^>_{\pm}|>1$ and $|\zeta^<_{\pm}|<1$ hold. Thus, from (\ref{eq:tilde_psi}), $\tilde\Psi\in W_{\lambda}\setminus \{\mathbf{0}\}$ satisfies $\sum_{x\in\mathbb{Z}}\|\tilde\Psi(x)\|_{\mathbb{C}^2}^2<\infty$ if and only if there exists $\varphi \in \mathbb{C}^2\setminus \{\mathbf{0}\}$ such that
\begin{align*}
\prod _{i=0}^{k-1} T_{i}^{+} T_{+} \varphi  & \in \ker\left(\tilde{T}_{k}^{+} -\zeta _{+}^{< }\right)\\
\Longleftrightarrow \varphi  & \in \ker\left(\left(\tilde{T}_{k}^{+} -\zeta _{+}^{< }\right)\prod _{i=0}^{k-1} T_{i}^{+} T_{+}\right) ,
\end{align*}
for all $ k\in \{0,\dotsc ,n_{+} -1\}$ and
\begin{align*}
\left(\prod _{i=k}^{n_{-} -1} T_{i}^{-}\right)^{-1} T_{-} \varphi  & \in \ker\left(\tilde{T}_{k}^{-} -\zeta _{-}^{ >}\right)\\
\Longleftrightarrow \varphi  & \in \ker\left(\left(\tilde{T}_{k}^{-} -\zeta _{-}^{ >}\right)\left(\prod _{i=k}^{n_{-} -1} T_{i}^{-}\right)^{-1} T_{-}\right)
\end{align*}
for all $ k\in \{0,\dotsc ,n_{-} -1\}$. Here $\iff$ denotes ``if and only if''. However, 
\[
\ker\left(\left(\prod _{i=0}^{n_{+} -1} T_{i}^{+} -\zeta _{+}^{< }\right) T_{+}\right) =\ker\left(\left(\tilde{T}_{k}^{+} -\zeta _{+}^{< }\right)\prod _{i=0}^{k-1} T_{i}^{+} T_{+}\right) ,
\]
holds for all $ k\in \{0,\dotsc ,n_{+} -1\}$ and 
\[
\ker\left(\left(\prod _{i=0}^{n_{-} -1} T_{i}^{-} -\zeta _{-}^{ >}\right) T_{-}\right) =\ker\left(\left(\tilde{T}_{k}^{-} -\zeta _{-}^{ >}\right) \ \left(\prod _{i=k}^{n_{-} -1} T_{i}^{-}\right)^{-1} T_{-}\right) ,\]
holds for all $ k\in \{0,\dotsc ,n_{-} -1\}$.
Thus, the conditions can be summarized in the second condition of the theorem. Therefore, the statement is proved.
\end{proof}
\end{theorem}
This theorem implies that the eigenvalue problem is to find the solution $\lambda\in[0,2\pi)$ of a single equation obtained from the second condition of the theorem in the range of $\left( A^{\pm }\right)^{2} -\left| \alpha _{0}^{\pm }\right| ^{2}\left| \alpha _{1}^{\pm }\right| ^{2} \cdots \left| \alpha _{n_{\pm } -1}^{\pm }\right| ^{2}  >0$. We can also see that if $e^{i\lambda} \in \sigma( U)$, the associated eigenvector $\Psi\in\ker(U-e^{i\lambda})\setminus\{\mathbf{0}\}$ is given as $\Psi=J^{-1}\tilde\Psi$ where 
\begin{align*}
\tilde{\Psi } (x) =\begin{cases}
\displaystyle(\zeta_{+}^<)^{m_{x}^{+}}\prod _{i=0}^{r_{x}^{+} -1} T_{i}^{+ } T_{+} \varphi , & x >x_{+} ,\\
\displaystyle\prod _{i=0}^{x-1} T_{i} \varphi , & 0< x\leq x_{+} ,\\
\varphi , & x=0,\\
\displaystyle\prod _{i=1}^{| x| } T_{-i}^{-1} \varphi , & x_{-} \leq x< 0,\\
\displaystyle(\zeta_{-}^>)^{-m_{x}^{-}} \ \left(\prod _{i=r_{x}^{-}}^{n_{-} -1} T_{i}^{- }\right)^{-1} T_{-} \varphi , & x< x_{-} ,
\end{cases}
\end{align*}
with $\displaystyle\varphi\in\ker\left(\left(\prod _{i=0}^{n_{+} -1} T_{i}^{+ } -\zeta _{+ }^{< }\right) T_{+}\right) \cap \ker\left(\left(\prod _{i=0}^{n_{-} -1} T_{i}^{- } -\zeta _{- }^{ >}\right) T_{-}\right).$

Furthermore, we can quantitatively evaluate localization by deriving time-averaged limit distribution defined by
\begin{equation*}
\overline{\nu }_{\infty } (x)=\lim _{T\rightarrow \infty }\frac{1}{T}\sum _{t=0}^{T-1}\left\Vert \left( U^{t} \Psi _{0}\right) (x)\right\Vert _{\mathbb{C}^{2}}^{2}.
\end{equation*}
The time-averaged limit distribution can be calculated by the eigenvectors of $U$ \cite{Segawa2016-qu}. For multiplicity $m_{\lambda } =\dim\ker (U-e^{i\lambda } )$ and complete orthonormal basis $\Psi _{j}^{\lambda } \in \ker (U-e^{i\lambda } ),\ j=1,2,\dotsc ,m_{\lambda }$, the following holds:
\begin{equation*}
\overline{\nu }_{\infty } (x)=\sum _{e^{i\lambda } \in \sigma (U)}\sum _{j,k=1}^{m_{\lambda }}\overline{\left< \Psi _{k}^{\lambda } ,\Psi _{0}\right> }\left< \Psi _{j}^{\lambda } ,\Psi _{0}\right> \left< \Psi _{k}^{\lambda } (x),\Psi _{j}^{\lambda } (x)\right>. 
\end{equation*}
Here, we show some important facts for calculating $\overline{\nu }_{\infty }$.
\begin{lemma}
\label{Lem:dimker}
$U$ has at most a finite number of eigenvalues with the multiplicity of 1, that is, 
\begin{equation*}
\dim\ker\left( U-e^{i\lambda }\right) =1\ 
\end{equation*}
and
\begin{equation*}
\sum _{e^{i\lambda } \in \sigma (U)}\dim\ker\left( U-e^{i\lambda }\right) < \infty .
\end{equation*}
\end{lemma}
\begin{proof}
By definition and Theorem \ref{Theorem Ker}, If $\dim\ker\left(\left(\prod\limits _{i=0}^{n_{+} -1} T_{i}^{+} -\zeta _{+}^{< }\right) T_{+}\right) =2$, then \\ $\left(\prod\limits _{i=0}^{n_{+} -1} T_{i}^{+} -\zeta _{+}^{< }\right) T_{+}$ becomes zero matrix and \ $\prod\limits _{i=0}^{n_{+} -1} T_{i}^{+} =\zeta _{+}^{< }$. This is clearly a contradiction. By similar discussion for $\left(\prod\limits _{i=0}^{n_{-} -1} T_{i}^{-} -\zeta _{-}^{ >}\right) T_{-} ,$ we have
\begin{equation*}
\dim\ker\left(\left(\prod _{i=0}^{n_{+} -1} T_{i}^{+} -\zeta _{+}^{< }\right) T_{+}\right) =\dim\ker\left(\left(\prod _{i=0}^{n_{-} -1} T_{i}^{-} -\zeta _{-}^{ >}\right) T_{-}\right) =1.
\end{equation*}
Also,
\begin{equation*}
\ker\left(\left(\prod _{i=0}^{n_{+} -1} T_{i}^{+} -\zeta _{+}^{< }\right) T_{+}\right) =\ker\left(\left(\prod _{i=0}^{n_{-} -1} T_{i}^{-} -\zeta _{-}^{ >}\right) T_{-}\right)
\end{equation*}
is immediately shown. Subsequently, we can see that
\begin{align*}
 & \dim\ker\left( U-e^{i\lambda }\right)\\
 & \ \ =\dim\left(\ker\left(\left(\prod _{i=0}^{n_{+} -1} T_{i}^{+} -\zeta _{+}^{< }\right) T_{+}\right) \cap \ker\left(\left(\prod _{i=0}^{n_{-} -1} T_{i}^{-} -\zeta _{-}^{ >}\right) T_{-}\right)\right) =1.\ 
\end{align*}
Furthermore, $e^{i\lambda } \in \sigma (U)$ has to be a root of the equation
\begin{equation*}
\operatorname{det}\left(\left(\prod _{i=0}^{n_{+} -1} T_{i}^{+} -\zeta _{+}^{< }\right) T_{+}\right) =0.
\end{equation*}
Hence, the number of $e^{i\lambda }$ satisfying the above equation is finite, and we complete the proof.
\end{proof}
Therefore, $\overline{\nu }_{\infty } (x)$ can be written as
\begin{align}
\label{eq:time_ave}
\overline{\nu }_{\infty } (x)=\sum _{e^{i\lambda } \in \sigma (U)}\left| \left< \Psi ^{\lambda } ,\Psi _{0}\right> \right| ^{2}\left\Vert \Psi ^{\lambda } (x)\right\Vert _{\mathbb{C}^{2}}^{2}.
\end{align}

\section{Results}
\label{sec:3}
\subsection{Homogeneously periodic model}

\begin{proposition}
\label{prop:homo}
Let $ n_{+} =n_{-} =n,\ x_{+} =x_{-} =0,\ C^{+}_{k} =C^{-}_{k} =C_{k},$ and $T^{+}_{k} =T^{-}_{k} =T_{k}$. The coin matrices become
\[
C_{x} =C_{r_{x}},
\]
 where $r_{x} =x\ (\bmod \ n)\in \{0,\dotsc ,n -1\}$. 
Then, the model does not exhibit localization, that is, $\sigma(U)=\emptyset$. 
\begin{proof}
Let $\zeta^{< }=\zeta _{+ }^{< }=\zeta _{- }^{< }$ and $\zeta^{> }=\zeta _{+ }^{> }=\zeta _{- }^{> }$. Then for all $\lambda$,
\[
\ker\left(\prod _{i=0}^{n-1} T_{i} -\zeta^{< }\right) \cap \ker\left(\prod _{i=0}^{n-1} T_{i} -\zeta^{ >}\right) =\{\mathbf{0}\}
\]
holds. Thus from Theorem \ref{Theorem Ker}, it is proved that $\sigma(U)=\emptyset$.
\end{proof}
\end{proposition}
From (\ref{eq:time_ave}), we can also see that the time-averaged limit distribution $\overline{\nu }_{\infty } (x)$ is always $0$ for all positions $x\in\mathbb{Z}$. As a remark, when $\alpha_x=0$ for some $x$ (excluded case), the model exhibits localization in the finite interval since the quantum walker is reflected at position $x$ where $\alpha_x=0$. This fact is described in detail by \cite{Shikano2010} as a lemma for the specific case
\[
C_x=\left[\begin{array}{cc}
\cos (2 \pi a x) & -\sin (2 \pi a x) \\
\sin (2 \pi a x) & \cos (2 \pi a x)
\end{array}\right],
\]
where $a$ is a rational number. With our Proposition \ref{prop:homo}, we can say that their lemma claims the only case in which the model exhibits localization. Our result contributes to solving one of the open problems mentioned in \cite{Shikano2010} for the general coin matrices. Moreover, note that when $n=2$, and if initial state $\Psi_0$ satisfies $\Psi_0(x)=\mathbf{0}$ for all odd positions $x$ or all even positions $x$, this model can be regarded as a time-dependent two-period quantum walks studied in \cite{Machida2010-kc}.
\subsection{Periodic model with one-defect}
\begin{figure}
\begin{subfigure}{0.49\textwidth}
\centering
\includegraphics[width=0.75\linewidth]{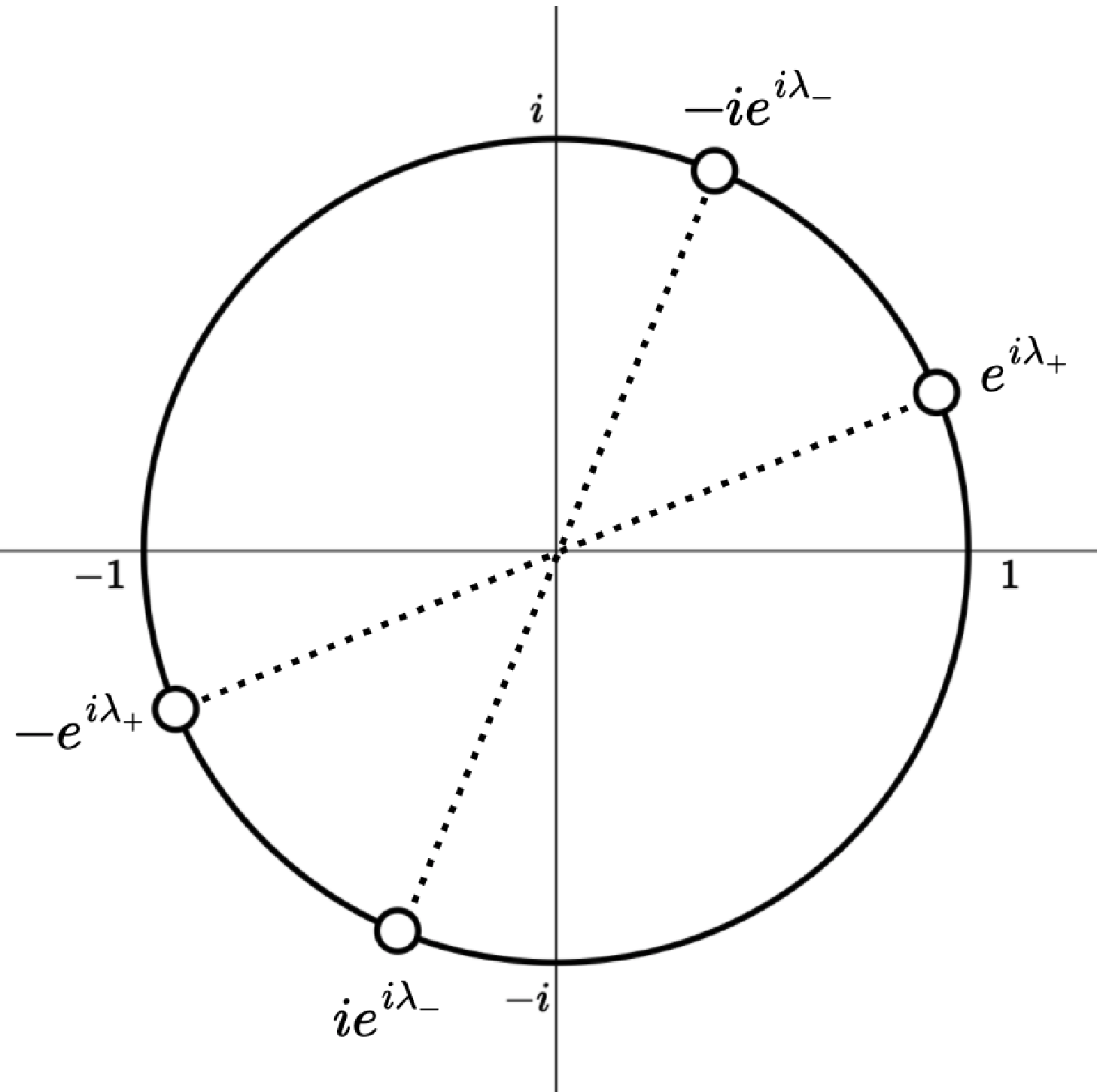} 
\caption{Eigenvalues of $U$}
\end{subfigure}
\begin{subfigure}{0.49\textwidth}
\centering
\includegraphics[width=0.95\linewidth]{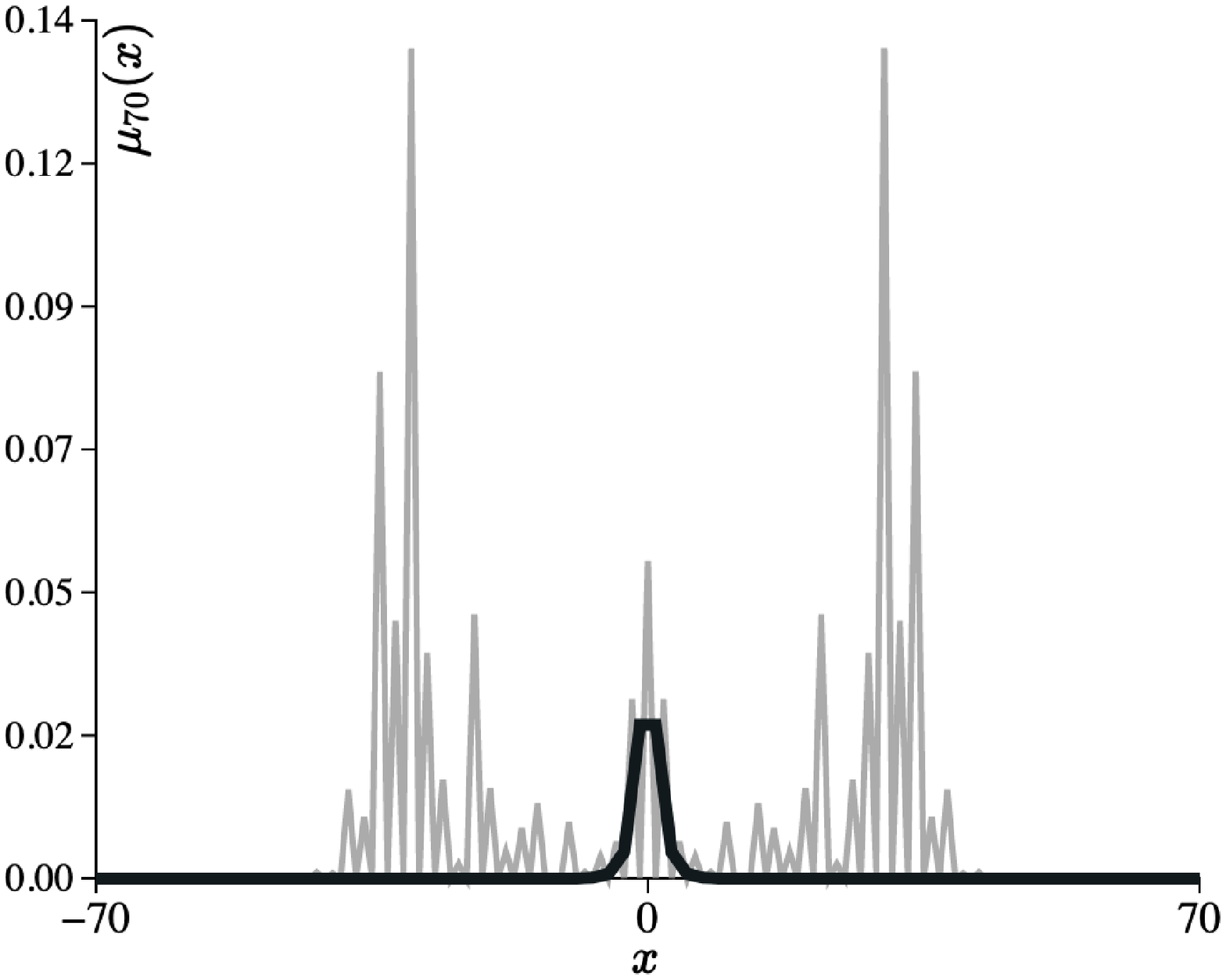}
\caption{Probability distribution}
\end{subfigure}
\caption{The example of Proposition \ref{prop:one} with parameters $\Delta _{1} =\frac{\pi }{2} ,\ \Delta _{2} =-\frac{\pi }{2} ,\ \alpha _{0} =1,\ \alpha _{1} =\alpha _{2} =\beta _{2} =\frac{1}{\sqrt{2}},\ \beta _{1} =\frac{i}{\sqrt{2}}$. (a) illustrates the eigenvalues of the time evolution operator. (b) shows the probability distribution at time $t=70$ (gray line) and the time-averaged limit distribution calculated from (\ref{eq:time_ave}) (bold black line). Here, the initial state is given as $\Psi_0(0)=[\frac{1}{\sqrt{2}},\frac{1}{\sqrt{2}}]^T$ and $\Psi_0(x)=\mathbf{0}$ for $x\neq 0$, where $T$ denotes the transpose operator}
\label{fig:1}  
\end{figure}

Next, we consider a one-defect model with a different coin matrix $C_0$ acts only on the origin in the homogeneously periodic model defined above. This is an extension of the usual one-defect model treated in previous study \cite{Kiumi2021-yg} by replacing coin matrices in positive and negative parts with periodic coins. The model is given by setting $n_+=n_-=n,\ x_{+} =1,\ x_{-} =0,\ C_{k}^{+ } =C_{k}^{- } =C_{k+1}$, and the transfer matrix can also be written as $T_{k}^{+ } =T_{k}^{- } =T_{k+1}$.
Then, coin matrices become
\[
C_{x} =\begin{cases}
C_{r^+_x+1 } , & x >0,\ \\
C_{0} , & x=0,\\
C_{r^-_{x} +1} , & x< 0.
\end{cases}\]
We get the analytical result for this model with $n=2$. From Theorem \ref{Theorem Ker}, $e^{i\lambda}\in\sigma(U)$ if and only if followings hold:
\begin{align*}
&( 1) \ \left(\cos( 2\lambda -\Delta _{1} -\Delta _{2}) +\Re \left( \beta _{1}\overline{\beta _{2}}\right)\right)^{2} -| \alpha _{1}| ^{2}| \alpha _{2}| ^{2}  >0,\\
&( 2) \ \ker\left(\left( T_{2} T_{1} -\zeta ^{< }\right) T_{0}\right) \cap \ker\left( T_{2} T_{1} -\zeta ^{ >}\right) \neq\{\mathbf{0}\},
\end{align*}
where $\Re$ denotes the real part of a complex number.
\begin{proposition}
\label{prop:one}
Let $\beta _{0} =0,\ | \beta _{1}| =| \beta _{2}| =| \beta | \neq 0,\ \Delta _{0} =\frac{\Delta _{1} +\Delta _{2} +\arg \beta _{1} -\arg \beta _{2} +\pi }{2}$. Then,
\[
\sigma( U) =\begin{cases}
\left\{\pm e^{i\lambda _{+}} ,\pm ie^{i\lambda _{-}}\right\} , & \Im \left( \beta _{1}\overline{\beta _{2}}\right) \  >0,\\
\left\{\pm e^{i\lambda _{-}} ,\pm ie^{i\lambda _{+}}\right\} , & \Im \left( \beta _{1}\overline{\beta _{2}}\right) \ < 0,\\
\left\{\pm e^{i\frac{\Delta _{1} +\Delta _{2}}{2}}\right\} , & \arg \beta _{1} =\arg \beta _{2} ,\\
\left\{\pm ie^{i\frac{\Delta _{1} +\Delta _{2}}{2}}\right\} , & \arg \beta _{1} =\arg \beta _{2} +\pi ,
\end{cases}
\]
where
\[
e^{i\lambda _{\pm }} =e^{i\frac{\Delta _{1} +\Delta _{2}}{2}}\left( \pm \sqrt{B_{+}} +i\sqrt{B_{-}}\right) ,\ B_{\pm } =\frac{| \beta | \pm \sqrt{| \beta | ^{2} -\Im ^{2}\left( \beta _{1}\overline{\beta _{2}}\right)}}{2| \beta | }.
\]
\end{proposition}
The example of Proposition \ref{prop:one} is shown in Figure \ref{fig:1}.
\subsection{Two-phase periodic model}
\begin{figure}
\begin{subfigure}{0.49\textwidth}
\centering
\includegraphics[width=0.75\linewidth]{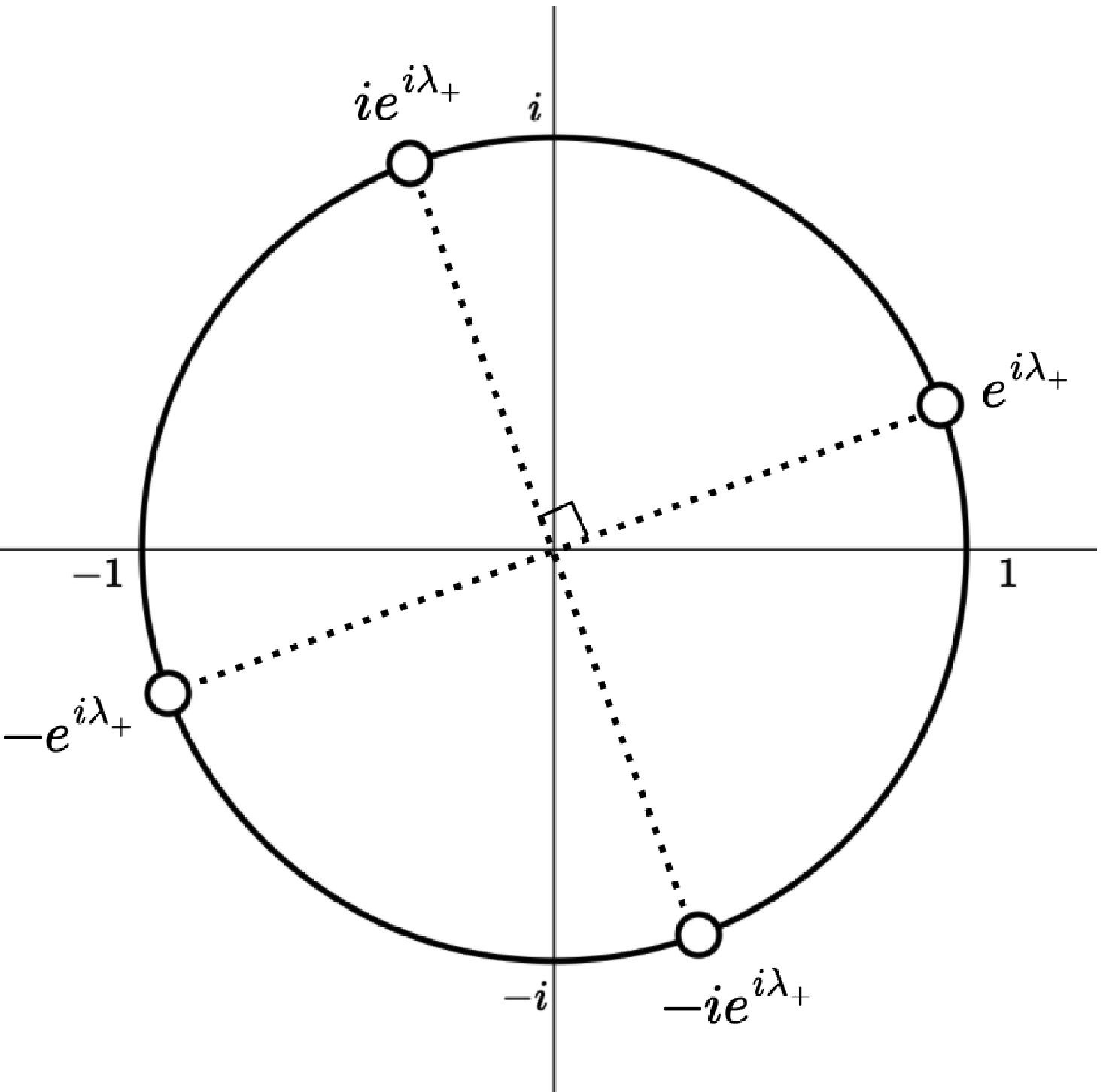} 
\caption{Eigenvalues of $U$}
\end{subfigure}
\begin{subfigure}{0.49\textwidth}
\centering
\includegraphics[width=0.95\linewidth]{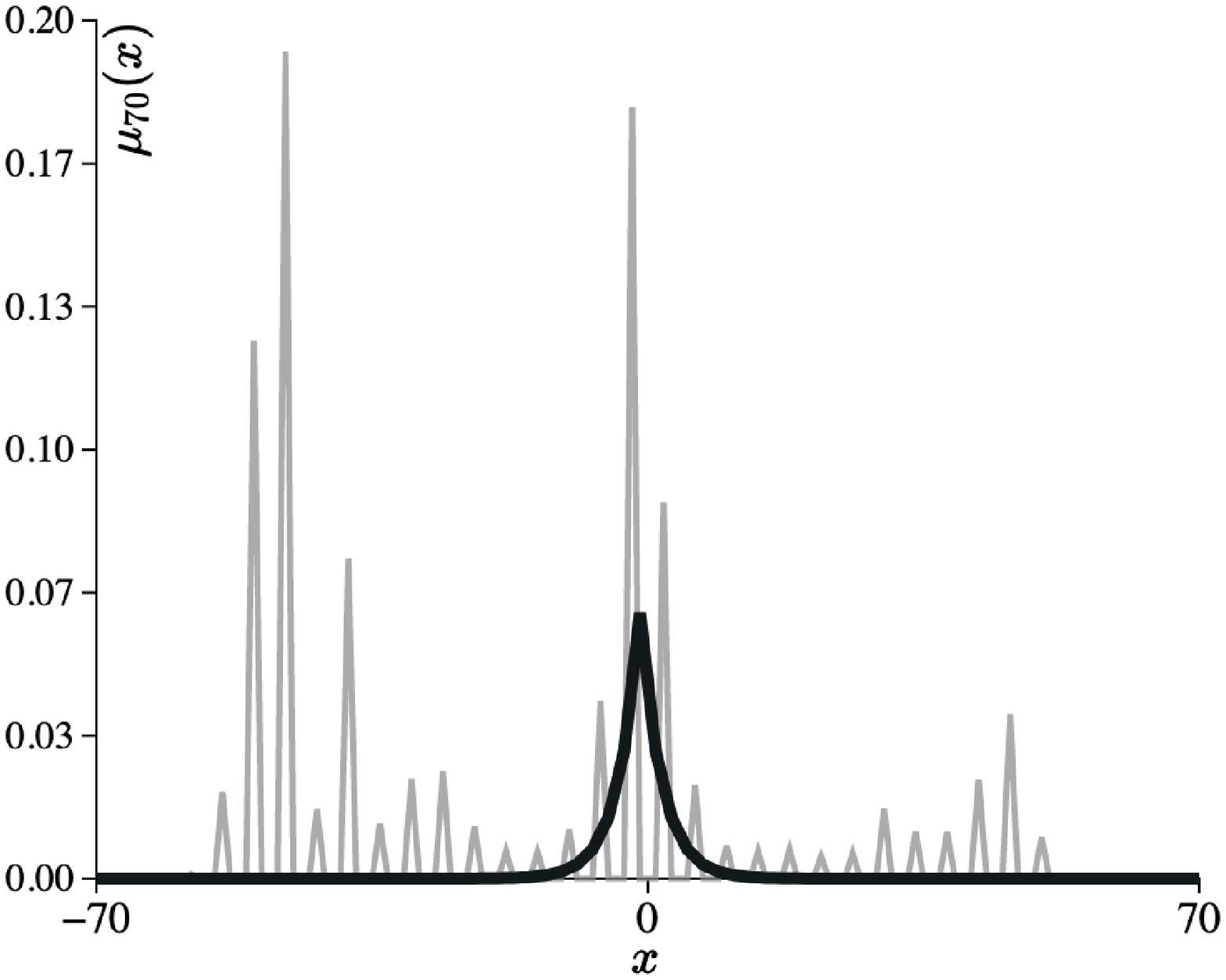}
\caption{Probability distribution}
\end{subfigure}
\caption{The example of Proposition \ref{prop:two1} with parameters $\Delta _{0} =\frac{\pi }{2} ,\ \Delta _{1} =-\frac{\pi }{2} ,\ \beta _{m} =\frac{1}{\sqrt{2}} e^{\frac{\pi }{4} i},\ \alpha _{p,0} =\alpha _{m,0} =\beta _{p} =\frac{1}{\sqrt{2}},\ \alpha _{p,1} =\alpha _{m,1} =1$. (a) illustrates the eigenvalues of the time evolution operator. (b) shows the probability distribution at time $t=70$ (gray line) and the time-averaged limit distribution calculated from (\ref{eq:time_ave}) (bold black line). Here, the initial state is given as $\Psi_0(0)=[\frac{1}{\sqrt{2}},\frac{1}{\sqrt{2}}]^T$ and $\Psi_0(x)=\mathbf{0}$ for $x\neq 0$}
\label{fig:2}  
\end{figure}

\begin{figure}
\begin{subfigure}{0.49\textwidth}
\centering
\includegraphics[width=0.75\linewidth]{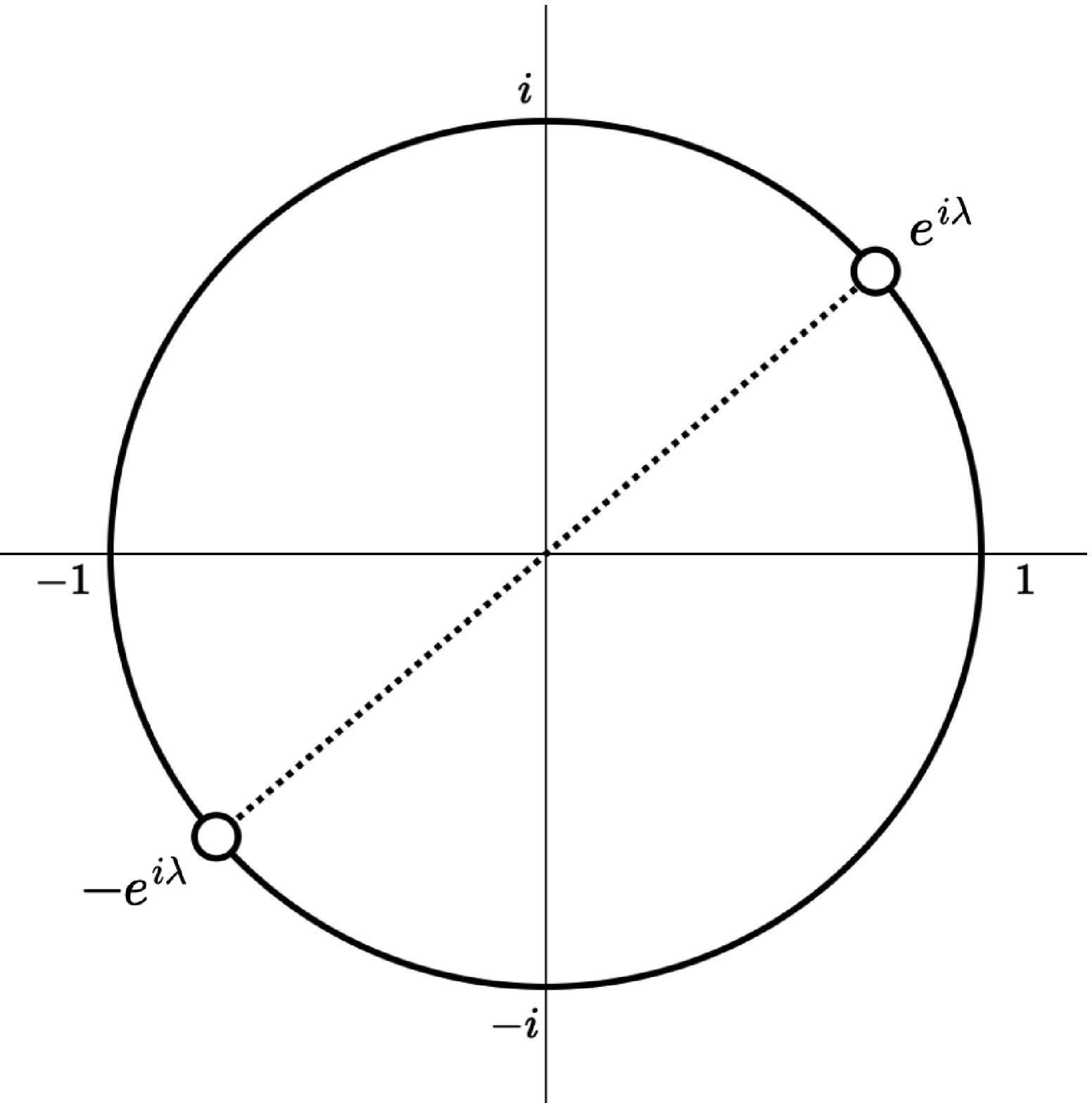} 
\caption{Eigenvalues of $U$}
\end{subfigure}
\begin{subfigure}{0.49\textwidth}
\centering
\includegraphics[width=0.95\linewidth]{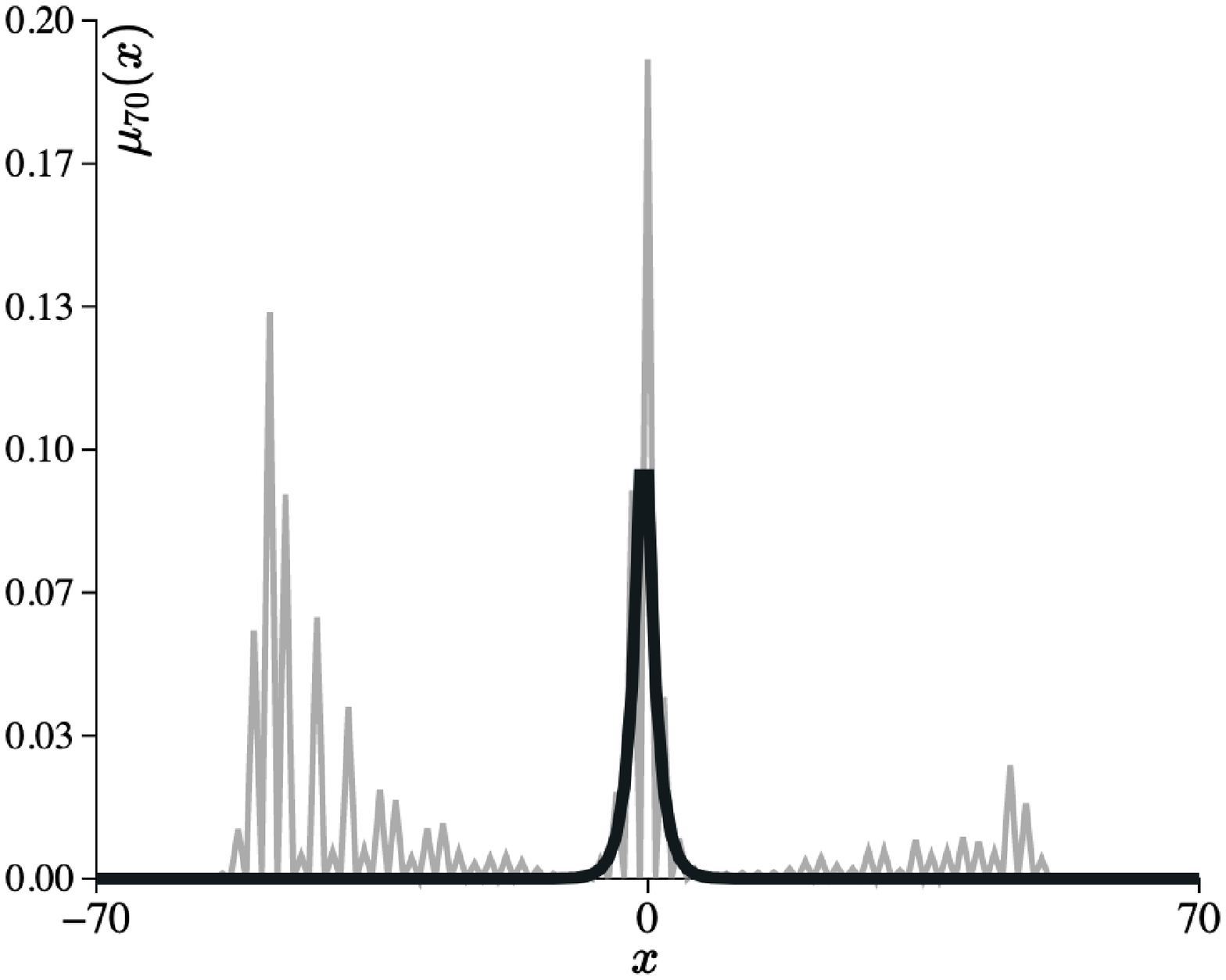}
\caption{Probability distribution}
\end{subfigure}
\caption{The example of Proposition \ref{prop:two2} with parameters $\Delta _{0} =\frac{\pi }{2} ,\ \Delta _{1} =-\frac{\pi }{2} ,\ \beta _{m} =\frac{1}{\sqrt{2}} e^{\frac{\pi }{4} i} ,\beta _{p} =\frac{1}{\sqrt{2}}$. (a) illustrates the eigenvalues of the time evolution operator. (b) shows the probability distribution at time $t=70$ (gray line) and the time-averaged limit distribution calculated from (\ref{eq:time_ave}) (bold black line). Here, the initial state is given as $\Psi_0(0)=[\frac{1}{\sqrt{2}},\frac{1}{\sqrt{2}}]^T$ and $\Psi_0(x)=\mathbf{0}$ for $x\neq 0$.}
\label{fig:3}  
\end{figure}

Next, we consider a two-phase model where two different groups of periodic coin matrices act on each of non-negative and negative parts, respectively. This is an extension of the usual two-phase model treated in previous study \cite{Kiumi2021-yg} by replacing coin matrices in each of non-negative and negative parts with periodic coins. The model is given by setting $n_+=n_-=n,\ x_{+} =0,\ x_{-} =0$. We write
\begin{align*}
 & \left( C_{k}^{- } ,\alpha _{k}^{- } ,\beta _{k}^{- } ,\Delta _{k}^{- } ,T_{k}^{- }\right) =( C_{m,k} ,\alpha _{m,k} ,\beta _{m,k} ,\Delta _{m,k} ,T_{m,k}),\\
 & \left( C_{k}^{+ } ,\alpha _{k}^{+ } ,\beta _{k}^{+ } ,\Delta _{k}^{+ } ,T_{k}^{+ }\right) =( C_{p,k} ,\alpha _{p,k} ,\beta _{p,k} ,\Delta _{p,k} ,T_{p,k})
\end{align*}
for $k=0,1\dots n-1$. Then, the coin matrices become
\[
C_{x} =\begin{cases}
C_{p,r_{x}} , & x\geq 0,\ \\
C_{m,r_{x}} , & x< 0,
\end{cases}
\]
where $r_{x} =x\ (\bmod \ n)\in \{0,\dotsc ,n -1\}$. We get analytical results for this model with $n=2$. From Theorem \ref{Theorem Ker}, $e^{i\lambda}\in\sigma(U)$ if and only if followings hold:
\begin{align*}
 & ( 1) \ \left(\cos( 2\lambda -\Delta _{m,0} -\Delta _{m,1}) +\Re \left( \beta _{m,0}\overline{\beta _{m,1}}\right)\right)^{2} -| \alpha _{m,0}| ^{2}| \alpha _{m,1}| ^{2}  >0,\\
 & \ \ \ \ \ \left(\cos( 2\lambda -\Delta _{p,0} -\Delta _{p,1}) +\Re \left( \beta _{p,0}\overline{\beta _{p,1}}\right)\right)^{2} -| \alpha _{p,0}| ^{2}| \alpha _{p,1}| ^{2}  >0,\\
 & ( 2) \ \ker\left( T_{p,1} T_{p,0} -\zeta _{p}^{< }\right) \cap \ker\left( T_{m,1} T_{m,0} -\zeta _{m}^{ >}\right) \neq\{\mathbf{0}\}.
\end{align*}
\begin{proposition}
\label{prop:two1}
Let $\displaystyle \beta _{m,1} =\beta _{p,1} =0,\ \beta _{m,0} =\beta _{m} ,\ \beta _{p,0} =\beta _{p} ,\ \Delta _{p,1} =\Delta _{m,1} =\Delta _{1} ,\ \Delta _{p,0} =\Delta _{m,0} =\Delta _{0}$, Then,
$\sigma( U) \neq \emptyset$ if and only if $\Re \left( \beta _{m}\overline{\beta _{p}}\right) < | \beta _{m}| ^{2}$ and $\Re \left( \beta _{m}\overline{\beta _{p}}\right) < | \beta _{p}| ^{2}$ and $\sigma(U)$ becomes
\[
\sigma( U) =\begin{cases}
\left\{\pm e^{i\lambda _{-}} ,\pm ie^{i\lambda _{-}}\right\}, & \mathfrak{I}\left( \beta _{m}\overline{\beta _{p}}\right) < 0,\ \\
\left\{\pm e^{i\lambda _{+}} ,\pm ie^{i\lambda _{+}}\right\}, & \mathfrak{I}\left( \beta _{m}\overline{\beta _{p}}\right)  >0,
\end{cases}
\]
where
\[
e^{i\lambda _{\pm }} =e^{i\frac{\Delta _{0} +\Delta _{1}}{2}}\left(\sqrt{B_{+}} \pm i\sqrt{B_{-}}\right) ,\ B_{\pm } =\frac{|\beta _{m} -\beta _{p} |\pm \sqrt{| \beta _{p} -\beta _{m}| ^{2} -\Im ^{2}\left( \beta _{m}\overline{\beta _{p}}\right)}}{2|\beta _{m} -\beta _{p} |}.
\]

\end{proposition}
The example of Proposition \ref{prop:two1} is shown in Figure \ref{fig:2}.
\begin{proposition}
\label{prop:two2}
Let $\displaystyle \Delta _{p,0} =\Delta _{m,0} =\Delta _{0} ,\ \Delta _{p,1} =\Delta _{m,1} =\Delta _{1} ,\ \beta _{m,0} =\beta _{m,1} =\beta _{m} ,\ \beta _{p,0} =\beta _{p,1} =\beta _{p}$, Then,
$\sigma( U) \neq \emptyset$ if and only if $\Re \left( \beta _{m}\overline{\beta _{p}}\right) < | \beta _{m}| ^{2}$ and $ \Re \left( \beta _{m}\overline{\beta _{p}}\right) < | \beta _{p}| ^{2}$. In this case,
\[
\sigma( U) =\left\{\pm e^{i\lambda }\right\},
\]
where
\[
e^{i\lambda } =e^{i\frac{\Delta _{0} +\Delta _{1}}{2}}\frac{\sqrt{| \beta _{p} -\beta _{m}| ^{2} -\Im ^{2}\left( \beta _{m}\overline{\beta _{p}}\right)} +i\mathfrak{I}\left( \beta _{m}\overline{\beta _{p}}\right)}{|\beta _{m} -\beta _{p} |}.\]

\end{proposition}
The example of Proposition \ref{prop:two2}, is shown in Figure \ref{fig:3}.

\section{Summary}
In previous studies \cite{Kiumi2021-yg,Kiumi2021-dp}, the eigenvalue analysis using a transfer matrix was performed for a model with homogeneous coin matrices in positions sufficiently far to the left and right, respectively, which include one-defect and two-phase models. This study focuses on the eigenvalue problem for a more generalized model in which the coin matrices are arranged periodically in positions sufficiently far to the left and right, respectively. Theorem \ref{Theorem Ker} is the main theorem, and it successfully provides the necessary and sufficient conditions for the eigenvalue problem. Furthermore, we showed in Lemma \ref{Lem:dimker} that the model has at most a finite number of eigenvalues with the multiplicity of 1, and we further discussed about the analytical formulation of the time-averaged limit distribution. Based on the main theorem, we considered the more specific models, which can be seen as generalizations of homogeneous, one-defect, and two-phase models. Proposition \ref{prop:homo} showed that if periodic coin matrices are arranged homogeneously, the model does not exhibit localization. Finally, Propositions \ref{prop:one}, \ref{prop:two1} and \ref{prop:two2} derived the concrete eigenvalues for specific models with periodicity 2.

For future research, further analysis using the transfer matrix for more general or different types of models, such as higher-dimensional and split-step \cite{Kitagawa2010-su} quantum walks, would be interesting. 
\section*{Acknowledgements}
\noindent
The author expresses sincere thanks and gratitude to Kei Saito for helpful comments and discussion.
\printbibliography


\end{document}